\documentclass[conference]{IEEEtran}
\IEEEoverridecommandlockouts
\usepackage{cite}
\usepackage{amsmath,amssymb,amsfonts}
\usepackage{algorithmic}
\usepackage{graphicx}
\usepackage{textcomp}
\usepackage{xcolor}
\def\BibTeX{{\rm B\kern-.05em{\sc i\kern-.025em b}\kern-.08em
    T\kern-.1667em\lower.7ex\hbox{E}\kern-.125emX}}
\begin{document}

\title{A General, Fault tolerant, Adaptive, Deadlock-free Routing Protocol for Network-on-chip
\thanks{P. Stroobant is funded by a Ph.D. grant of Ghent University, Special Research Fund (BOF). S. Abadal and E. Alarc\'on gratefully acknowledge support by the European Commision under grant H2020-FETOPEN-736876 (VISORSURF).}
}

\author{
\IEEEauthorblockN{Pieter Stroobant}
\IEEEauthorblockA{\textit{IDLab} \\
\textit{Ghent University - imec}\\
Ghent, Belgium \\
pieter.stroobant@ugent.be}
\and
\IEEEauthorblockN{Sergi Abadal}
\IEEEauthorblockA{\textit{N3Cat} \\
\textit{Universitat Polit\`ecnica de Catalunya}\\
Barcelona, Spain \\
abadal@ac.upc.edu}
\and
\IEEEauthorblockN{Wouter Tavernier}
\IEEEauthorblockA{\textit{IDLab} \\
\textit{Ghent University - imec}\\
Ghent, Belgium \\
wouter.tavernier@ugent.be}
\and
\IEEEauthorblockN{Eduard Alarc\'on}
\IEEEauthorblockA{\textit{N3Cat} \\
\textit{Universitat Polit\`ecnica de Catalunya}\\
Barcelona, Spain \\
eduard.alarcon@upc.edu}\\
\and
\IEEEauthorblockN{Didier Colle}
\IEEEauthorblockA{\textit{IDLab} \\
\textit{Ghent University - imec}\\
Ghent, Belgium \\
didier.colle@ugent.be}
\and
\IEEEauthorblockN{Mario Pickavet}
\IEEEauthorblockA{\textit{IDLab} \\
\textit{Ghent University - imec}\\
Ghent, Belgium \\
mario.pickavet@ugent.be}
}

\IEEEoverridecommandlockouts
\IEEEpubid{\makebox[\columnwidth][c]{978-1-5386-8552-5/18/\$31.00~\copyright2018 IEEE \hfill}
\hspace{\columnsep}\makebox[\columnwidth]{ }}

\maketitle

\IEEEpubidadjcol

\begin{abstract}
The paper presents a topology-agnostic greedy protocol for network-on-chip routing. The proposed routing algorithm can tolerate any number of permanent faults, and is proven to be deadlock-free. We introduce a specialized variant of the algorithm, which is optimized for 2D mesh networks, both flat and wireless. The adaptiveness and minimality of several variants this algorithm are analyzed through graph-based simulations.
\end{abstract}


\section{Introduction}

The ongoing downscaling of transitor technology has allowed to integrate increasingly large numbers of electronic systems in a single chip (i.e. system-on-chip). Network-on-chip (NoC) provides a communication system between the components of such a system-on-chip and has been shown to be a more scalable means of communication than traditional bus-based connections \cite{Dally2001}.

However, these small-scale technologies are expected to be increasingly prone to hardware defects \cite{Constantinides2007}. Also, the use of network-on-chip technology has been proposed for challenging applications such as software-defined metamaterials, which are deployed in extreme environments in which component failures are more likely to occur \cite{Abadal2017}. This creates a need for network-on-chip routing algorithms which are able to deal with such defects.

Additionally, a variety of NoC architectures have been proposed \cite{Erika2012}. The heterogenity in network topology is increased even further by power-gating techniques \cite{Chen2012} and the introduction of wireless, reconfigurable links \cite{Murray2014}. Routing algorithms should be able to handle a variety of network topologies.

The contribution of our paper is to propose and evaluate a routing protocol with the following properties:
\begin{itemize}
  \item Our algorithm achieves \emph{full fault coverage}: whenever a path between a transmitting node and the destination exist, a route will be found. This is in contrast to some methods (see Section \ref{sec:rel_work}), which require to deactivate healthy nodes that are part of a `fault region'.
  \item The presented routing scheme is \emph{topology-agnostic}, i.e. it works (in principle) on any topology. However, increasingly complex topologies require longer address sizes. Network layouts with a high regularity allow for a more efficient address representation (see Section \ref{ssec:addressing}).
  \item \emph{No routing table} is required by the algorithm. Thus, no hardware for such tables must be provided, which limits the area overhead of an implementation.
  \item The algorithm is \emph{fully distributed}: there is no single point of failure in the network 
  \item The proposed method is \emph{deadlock free}, and \emph{no virtual channels} are required
  \item When several paths between the source and destination nodes are possible, the routing method typically allows forwarding nodes some choice in the selection of the next hop. Combining this freedom with local congestion information may allow routers to adapt the route in order to aim at a lower congestion rate. This \emph{adaptiveness} is analysed in Section \ref{sec:evaluation}. However, in use cases which require that all packets arrive in order, the algorithm can be made deterministic as well.
  \item Whilst in operation, some newly occurring component (node or link) failures do not require to reconfigure the routing algorithm. This property can be leveraged to achieve \emph{no reconfiguration overhead when turning off parts of the chip to save power using power-gating techniques \cite{Chen2012}}.
  \item There is \emph{no guaranteed minimal path length}, that is, the length of some of the paths generated by the algorithm may suboptimal. Longer paths result in an increased latency, and thus should be avoided. Section \ref{sec:evaluation} evaluates the performance of the algorithm in terms of path length.
\end{itemize}

To the best of our knowledge, our work is the first topology-agnostic routing algorithm which provides full fault coverage without requiring the area overhead that comes with routing tables. This comes from the fact that we are the first ones to propose the use of \emph{spanning trees for greedy routing} in network-on-chip.

Our work is structured as follows: an overview of related work on fault tolerant routing techniques for networks-on-chip is given in Section \ref{sec:rel_work}. Subsequently, Section \ref{sec:DF_multi_georouting} explains the routing technique in its most general formulation, and provides proofs and insights in the properties of the routing method. Next, Section \ref{sec:DF_multi_georouting_grids} argues how to optimize this algorithm for (hierarchical) grids. After this, Section \ref{sec:evaluation} presents some key metrics for the proposed algorithm in a 2D mesh network setting. Finally, Section \ref{sec:conclusion} summarizes our work.

\section{Related Work} \label{sec:rel_work}
An overview of fault tolerant routing techniques in the context of network-on-chip is provided by \cite{Radetzki2013}. In the context of our work, we provide a short overview of the main techniques which offer online fault tolerance, i.e. the techniques which are able to deal with failures which occur after a chip has left the factory. We present these techniques in order of increasing complexity.

A first category of algorithms employ face routing based techniques to find paths around areas with faulty components. Maze routing is an important example of these techniques \cite{Fattah2015}. An important advantage of such methods is that recent information on the status of each link/node is required only by its neighbouring nodes, thus resulting in a low reconfiguration overhead. Furthermore, these methods provide full fault coverage: a route will be found whenever a path from a source node to its destination exists. The main disadvantage is that these methods are limited to planar graph topologies and cannot easily be extended to other topologies, such as 3D NoCs, wireless NoCs or a torus topology. Additionally, when a packet encounters an area with faulty components, the lack of information on the shape of this area may result in the message being routed along the border in a suboptimal direction, thus significantly increasing the path length.

Another important technique uses fault regions \cite{Fukushima2009}. This method applies to n-dimensional mesh networks, in which one or multiple faults are grouped in one or more rectangular non-overlapping regions. When encountering such a region, the protocol routes along the border of the fault region. The main issue with this technique is that the rectangular fault regions typically contain some healthy nodes as well. The healthy nodes on the inside of a fault region such a cannot receive messages and thus must be turned off, which results in a bad resource utilization.

Other authors extended this method by proposing to combine the fault regions with extra information regarding the connectivity of the neighbourhood of a router \cite{Jovanovic2009}. As routers take into account component information of nodes at an increasing hop distance, fault regions of increasingly complicated shapes are allowed. However, increasing the hop distance also results in a more complex forwarding process and exchanging this information increases the communication overhead.

As the hop distance within which local information is exchanged increases, nodes start to exchange information on the full graph topology. Such methods guarantee full fault coverage without making any assumptions on the graph structure. Hence this category of algorithms is topology-agnostic.
Sharing this connectivity information results in a large communication overhead and long reconfiguration times. Additionally, the routing information is usually stored in a routing table, increasing the hardware complexity of the routing module.
Packet switching routing algorithms, such as link state routing and distance vector routing, are part of this category.

A subcategory of routing schemes in which the full graph topology is processed employs spanning trees to ensure that the routing process remains deadlock free. An overview of these methods is given by \cite{Flich2012}. The fundamental algorithm in this category is up*/down* routing \cite{Schroeder1991}. In this method, a spanning tree is constructed over the network, and the `up' or `down' direction is assigned to each arc in the network (including the ones that are not part of the spanning tree). The directions are assigned in such a way that there is no cycle containing exclusively `up' or exclusively `down' arcs. To guarantee deadlock freeness, shortest paths between a source and destination vertex are searched under the restriction that no up edge can follow after a down edge (i.e. the `down-up' turn is forbidden). These path constraints were relaxed in \cite{Jouraku2007}, which uses uses a more fine grained categorization of edges, in which only few specific turns are prohibited.

Also related to this are several publications on deadlock free routing using multiple up*/down* trees \cite{Lysne2006, Flich2002, Ingebjorg2003}. Each of the trees has its own separate root, and corresponds a so-called `network-layer'. These methods require virtual channels: since routing in each of the trees is deadlock free, the multi-tree algorithm is guaranteed to be deadlock free when no more trees are concurrently in use than the number of available virtual channels.

\section{Deadlock free multi-tree geometric routing} \label{sec:DF_multi_georouting}

This section explains the workings of our deadlock free multi-tree geometric routing in a general network setting.

\subsection{Single-tree geometric routing} \label{ssec:single_tree}
Geometric routing protocols operate by assigning coordinates to each of the nodes in a network, and require that sending nodes specify the coordinates of the receiver in their transmitted messages. These coordinates are subsequently used in the forwarding process to route packets to their destination without requiring a route discovery step. When routing in a 2D NoC mesh network, including the physical coordinates (i.e. the position of a node in the raster) is a common idea. NoC algorithms such as dynamic XY routing \cite{Li2006} and the (odd-even) turn model \cite{Glass1992, Chiu2000} forward packets towards the neighbour for which the Manhattan distance to the destination is minimal, whilst taking into account some additional restrictions to guarantee the deadlock freeness of the routing scheme.

Unfortunately, greedy based routing using the Manhattan distance between the physical raster coordinates of nodes may, in the presence of faults, suffer from so called `local minima', i.e. nodes for which the Manhattan distance between each of their neighbours and the destination exceeds the Manhattan distance between themselves and the target location. Such local minima may cause packets to get stuck. Greedy embeddings work around this problem by assigning to each node a \emph{virtual coordinate}, over which a distance function is defined in such a way that no local minima exist.

\begin{figure}[ht]
    \centering
    \includegraphics[width=0.3\textwidth]{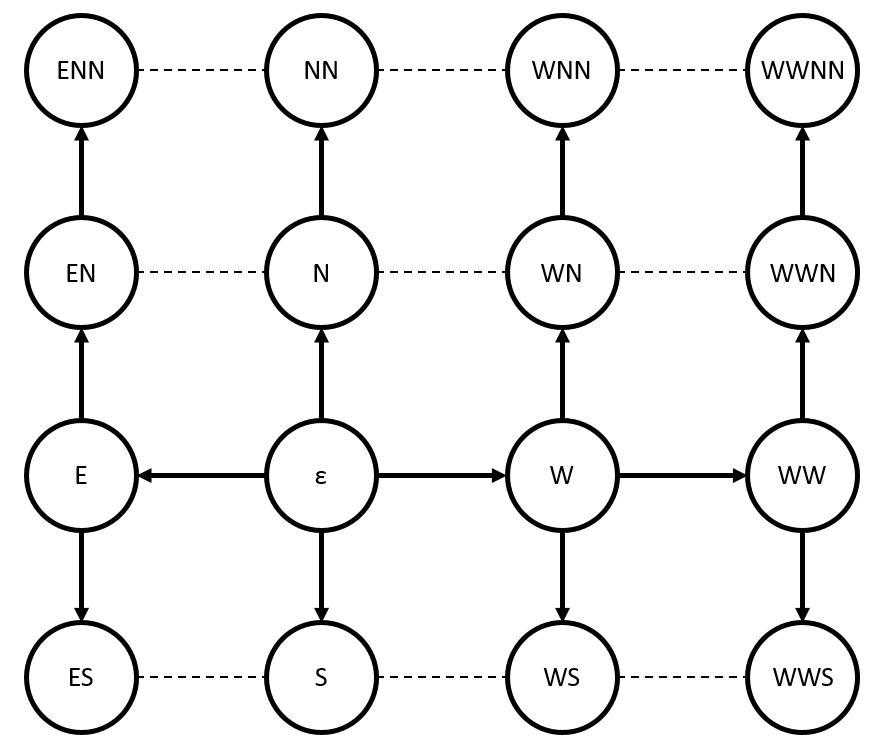}
    \caption{Example spanning tree for a fault-free 4x4 mesh network. The identifiers assigned to the arcs correspond to their compass direction.}
    \label{fig:4mesh_vertical}
\end{figure}

Our algorithm uses a tree-based embedding, in which coordinates are assigned based on a breadth-first search spanning tree, which consists of a subset of all bidirectional links in the network. The address of a node uniquely identifies the position of a node in the tree: each node assigns to each of its outgoing arcs a locally unique identifier, and the address of a node $n$ corresponds to the identifiers of the edges on the path between the root and $n$. As an example, a spanning tree for a fault-free 4x4 mesh and the corresponding tree coordinates are illustrated in Fig. \ref{fig:4mesh_vertical}.

Algorithms to set up such a spanning tree in a fully distributed manner have been studied in the literature. In \cite{Sahhaf2013}, a distributed spanning tree algorithm is presented in which all communication is between neighbouring nodes. At any moment during the algorithm, nodes only need to store the id corresponding to the root of the tree, and their own address within that tree. The algorithm guarantees that the node with the highest id becomes the root of the tree. Since the location of the root node is preferably as central as possible, the node ids could be preprogrammed such that centrally located nodes have higher ids.

These addresses allow to calculate the distance between any pair of nodes $(n_1, n_2)$ when travelling along the tree: starting from $n_1$, such a path travels upwards along the tree to the last shared ancestor of the two nodes and then heads downwards towards $n_2$. Assume that the address of $n_1$ is $(e_{1, 1}, e_{1, 2}, ..., e_{1, I})$, the address of $n_2$ is $(e_{2, 1}, e_{2, 2}, ..., e_{2, J})$ and the last shared ancestor of $n_1$ and $n_2$ is reached via the last shared edge $e_{1, K}=e_{2, K}$ of the two addresses. In this case the length of the path between $n_1$ and $n_2$ equals $I+J-2K$. Thus, the `tree distance' between any pair of nodes can be calculated by searching for the position of the last shared edge in their addresses.

\begin{figure}[ht]
    \centering
    \includegraphics[width=0.3\textwidth]{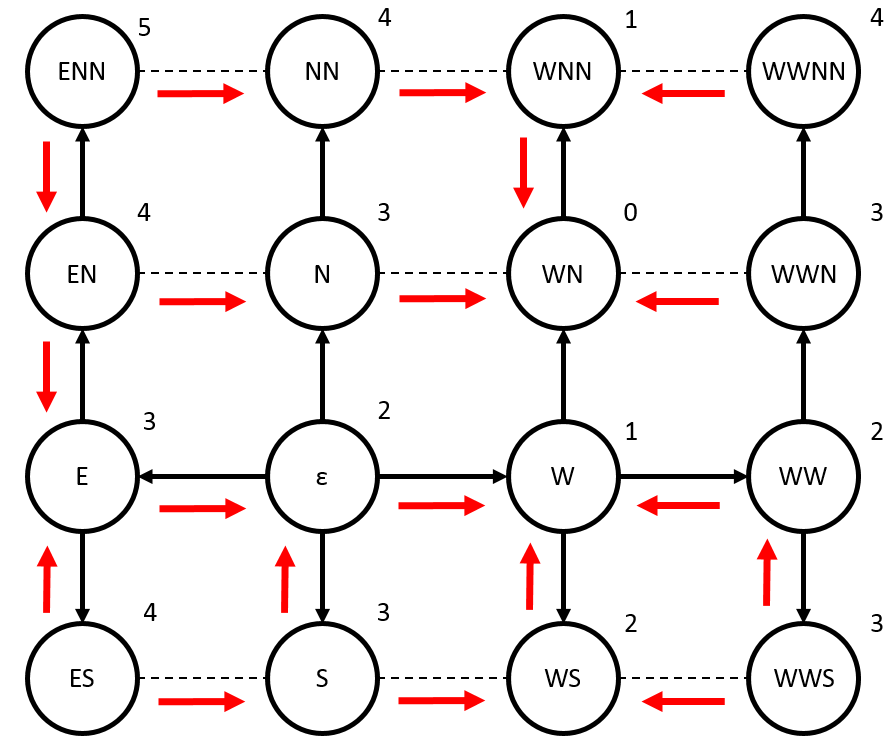}
    \caption{Example spanning tree and forwarding directions for routing towards the `WN' node in a fault-free 4x4 mesh network}
    \label{fig:4mesh_vertical_NW}
\end{figure}

Fig. \ref{fig:4mesh_vertical_NW} shows, for the network of Fig. \ref{fig:4mesh_vertical}, the distance between each node and the node with address WN. For each node, all possible neighbours to which a message addressed to node WN may be forwarded are indicated. Note that greedy routing is different from routing along the tree: greedy routing allows to use connections which are not part of the tree if they decrease the distance towards the destination.

\subsection{Deadlock freeness}
\label{ssec:deadlock_freeness}
In order to guarantee the deadlock freeness of the routing algorithm, some additional forwarding restrictions/rules must be taken into account in each node through which a packet passes.
Our approach is as follows: each bidirectional connection (edge) in the network is assumed to be implemented as two separate channels (arcs) with opposite directions. We categorize each channel according to the depth of its start and end node within the spanning tree and subsequently, deadlocks are avoided by placing restrictions on which arc types can be taken in each of the nodes.

Consider a general network topology over which a spanning tree has been constructed. Each node $n$ in this network has a depth $d(n)$ corresponding to the length of the path between the root of the tree and $n$ when travelling along the tree. Based on this depth, a connection starting at a node $n_1$ and ending at a node $n_2$ must be of one of the following types:
\begin{itemize}
  \item $d(n_2)<d(n_1)$: the connection is upwards ($\uparrow$)
  \item $d(n_2)=d(n_1)$: the connection is sideways ($\rightarrow$)
  \item $d(n_2)>d(n_1)$: the connection is downwards ($\downarrow$)
\end{itemize}



Deadlock freeness is achieved by restricting each path such that it consists of an initial sequence of edges which are either upwards or sideways, followed by a sequence of edges that are all downwards. Otherwise stated: for any path, the sequence of path types corresponding to the edges along the path is a string of the format: $(\uparrow | \rightarrow)* \downarrow*$.

Deadlock may occur only if there exists a cycle in the graph such that for any two subsequent arcs $a_1$ and $a_2$ in the cycle, we have that $a_1$ is dependent on $a_2$, i.e. any two subsequent arcs in the cycle must be part of some path along which a message may be routed.

Any cycle which contains an upwards arc must also contain a downwards arc (and vice versa). Thus, at some position in such a cycle a downwards channel must be followed by either a sideways or an upwards arc. This however implies that some packet must travel along a downward arc, followed by a sideways or an upwards arc, which is in contradiction with our restriction on the shape of a path. Hence, no cycle can contain any upwards or downwards arcs. This is similar to the reason why there are no cycles in up*/down* routing.

The only remaining cycles which may cause a deadlock are composed exclusively of sideways connections. However, for a packet which is routed over two sideways connections we consider the corresponding nodes $n_1$, $n_2$ and $n_3$ through which the packet passes. Since the distance between these nodes and the target of the packet decreases, but the length of the addresses of the nodes remains the same, the depth of the last shared ancestor between each of the nodes and the target node must increase. Also, the last shared ancestor between $n_1$ and $n_2$ is at the same depth as the last shared ancestor between $n_1$ and the target of the packet and similarly, the last shared ancestor between $n_2$ and $n_3$ is at the same depth as the last shared ancestor between $n_2$ and the target of the packet. Hence, the depth of the last shared ancestor between $n_2$ and $n_3$ exceeds the depth of the last shared ancestor between $n_1$ and $n_2$. However, the nodes $n_2$ and $n_3$ are once again the first nodes of another sequence of two sideways connections, where the depth of the last shared ancestor between the last pair of nodes will exceed the depth of the last shared ancestor between $n_2$ and $n_3$. Hence, the depth of the last shared ancestor between any two subsequent nodes keeps increasing whilst travelling along the cycle. However, since this is a cycle, the same two subsequent nodes will eventually occur again, which is in contradiction with the increasing depth of the last shared ancestor.

We conclude that enforcing that paths consisting of an initial sequence of edges which are either upwards or sideways, followed by a sequence of edges that are all downwards are indeed deadlock free. This restriction is easily achieved by allowing the use of downwards arcs only if the node at the end of the arc is an ancestor of the target node.

\subsection{Fault tolerance} \label{ssec:fault_tolerance}
Tree-based embeddings have some intrinsic properties which are advantageous when dealing with run-time faults.
Regarding link faults, reconfiguration is necessary only when a connection breaks which is part of the spanning tree. Also, non-tree links may remain in use even if one channel of the link fails, turning it into an unidirectional connection.
Node failures typically require to reconfigure the tree, unless the failing node is a leaf node (since in that case no part of the tree becomes disconnected). This property may also be used in power-gating techniques \cite{Chen2012}: leaf nodes or even entire subtrees can be switched off without requiring to reconfigure the algorithm.

\subsection{Multi-tree geometric routing} \label{ssec:multi_tree}
Single tree geometric routing may suffer in some cases from a discrepancy between the connectivity of nodes in the spanning tree and the connectivity of nodes in the full topology graph. This may cause a large tree distance between nodes which are physically located close to each other. This in turn may lead to suboptimal pathfinding. For example, in Fig. \ref{fig:4mesh_vertical}, when routing a packet from NN to WWNN it will be routed via the nodes with coordinates N, $\epsilon$, W, WW, WWN and WWNN. (Note that the link between NN and WNN and between N and WN may not be used due to the deadlock avoidance rule.)

\begin{figure}[ht]
    \centering
    \includegraphics[width=0.3\textwidth]{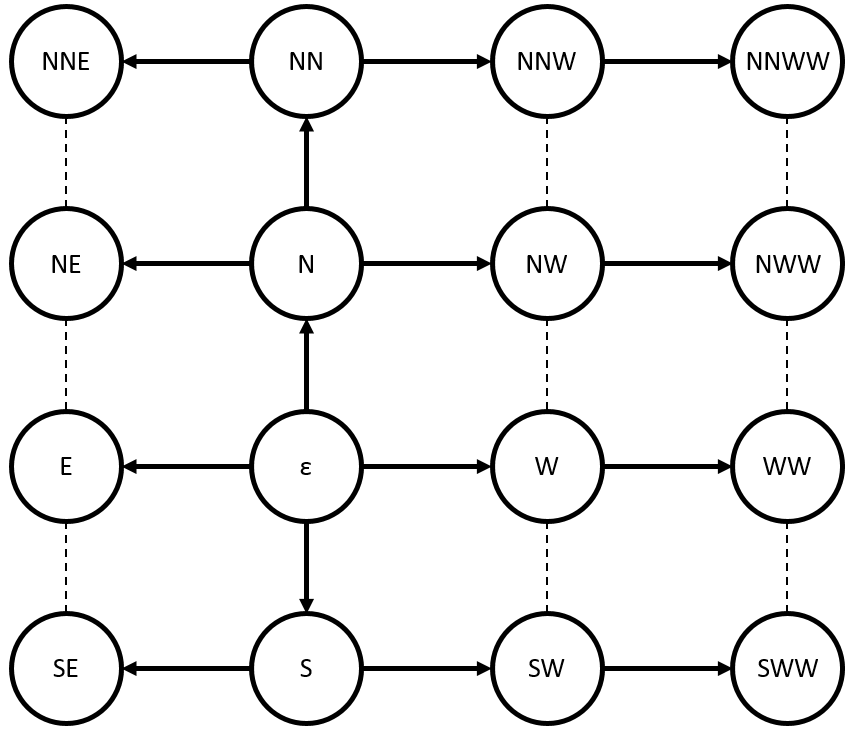}
    \caption{Example spanning tree for a fault-free 4x4 mesh network. The identifiers assigned to the arcs correspond to their compass direction.}
    \label{fig:4mesh_horizontal}
\end{figure}

This issue may be alleviated by using multiple spanning breadth-first trees for addressing. These trees should be constructed in such a way that nodes that are physically located close to each other but far from each other in one tree are close to each other in the other one. Fig. \ref{fig:4mesh_horizontal} shows an alternative tree for a fault free 4x4 network. Note that the NN and NNWW nodes, which were at a tree distance of 6 in Fig. \ref{fig:4mesh_vertical} are now at a tree distance of 2.

In this multi-tree algorithm, a packet is forwarded towards the neighbour for which the tree distance from the destination in any of the trees is minimized.

In order to guarantee deadlock freeness, all trees must be rooted in the same node: since these spanning trees are breadth-first trees, this ensures that the depth of a node in any of the trees is the same. The same path restriction rule that was introduced in subsection \ref{ssec:deadlock_freeness} can thus be applied in the multi-tree case. Similarly as before, one can argue that cycles cannot contain any upwards or a downwards arcs. Unfortunately, this is not the case for the argument regarding cycles consisting of sideways arcs, which holds only when routing in a single tree. We work around this issue by modifying the distance function, such that for sideways links the distance is always calculated over the same tree.

Regarding fault tolerance, it is worthwhile to notice that the multi-tree algorithm can also be successfully combined with power-gating techniques: nodes which are at the maximum depth are always leaf nodes in both trees and can be switched off without requiring to reconfigure the algorithm.

\section{Deadlock free multi-tree geometric routing in (hierarchical) grid networks} \label{sec:DF_multi_georouting_grids}
Grid based topologies are one of the most common network-on-chip topologies \cite{Bjerregaard2006}. This section discusses some optimizations of the previously discussed multi-tree routing algorithm for these (hierarchical) grids.

\subsection{Neighbour selection heuristic} \label{ssec:edge_sel_heur}
Multi-tree routers operate by randomly forwarding a message towards any of the nodes which are at a minimal tree distance from the destination. In the case of mesh networks, an additional heuristic is applied: for all neighbours which are at a minimal tree distance from the destination, the Manhattan distance to the destination is calculated as well. Of the neighbours with a minimal tree distance, a random node for which the Manhattan distance is minimized is selected.

\subsection{Tree growing algorithm} \label{ssec:tree_growing}
The multi-tree algorithms require to construct multiple breadth-first spanning trees rooted in the same node, such that nodes which are physically located in proximity to each other also have a low tree distance. We aim to achieve this target using direction-preferential tree growth: whenever a node at depth $d$ is connected to multiple nodes at depth $d-1$, the parent is picked based on the direction of the edge from that parent.
Fig. \ref{fig:4mesh_vertical} presents such a tree for a 4x4 mesh, with a preference for north-south connections over east-west connection. Fig. \ref{fig:4mesh_horizontal} shows a tree with a preference for connections in the east-west direction.


\subsection{Address compression} \label{ssec:addressing}
In wormhole routing, each packet is broken into several flits (flow control units). The header flit is processed by the router and should contain all routing information. When using multiple trees, the address of the target in each of the trees should be stored in the header flit. Using a full path representation (e.g. 2 bits per edge in a 2-dimensional flat grid) may easily result in address sizes which exceed the length of a flit, especially since introducing errors might increase the network radius, resulting in longer addresses.

This issue can be mitigated by exploiting the highly regular tree structure which is typical for direction-preferential trees. The addresses assigned by these trees typically contain many consecutive occurrences of arcs in the same direction. This allows to apply run-length encoding, in which each sequence of edges in the same direction is replaced by the direction of the edge and the number of occurrences. For example, an address NNNNEEEEEEN would be encoded as N4E6N1.

\section{Performance Evaluation}  \label{sec:evaluation}
This section employs graph-based simulations of 2D mesh network topologies to evaluate our algorithm with regard to several quality metrics. Two variants of the proposed deadlock free greedy routing algorithm were studied: one variant using a single direction-preferential tree, and one variant using two of these trees. Both variants their performance was evaluated both in 4x4 and 8x8 mesh networks for a varying probability of random link failures. In all subsequent figures, every measurement point is calculated over at least 250 000 source-destination pairs.

\begin{figure}[ht]
    \centering
    \includegraphics[width=0.41\textwidth]{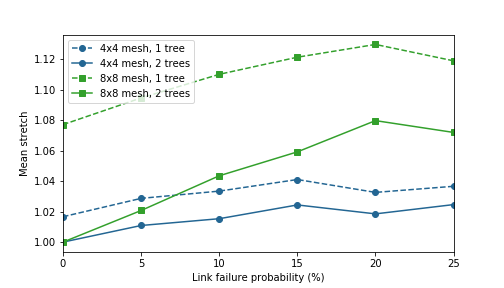}
    \caption{Mean stretch for routing using 1 or 2 trees, in a 4x4 or 8x8 mesh network}
    \label{fig:mean_stretch}
\end{figure}

Firstly, the quality of the routes was evaluated. For this purpose the mean stretch was calculated: for any source-destination query, the stretch equals the average path length of the routes provided by the algorithm divided by the length of the shortest path between the source and destination.
The mean stretch is visualised in Fig. \ref{fig:mean_stretch}. All algorithms provide low average stretch values (below 1.14). Unsurprisingly, the multi-tree algorithms outperform the single tree algorithms in all cases. Note that in the case of no failures, the multi-tree algorithm always returns a minimal route. We also observe that routing in the less complex 4x4 mesh results in smaller stretch values than routing within the larger 8x8 mesh. Increasing the link failure probability results in an increased mean stretch.

\begin{figure}[ht]
    \centering
    \includegraphics[width=0.41\textwidth]{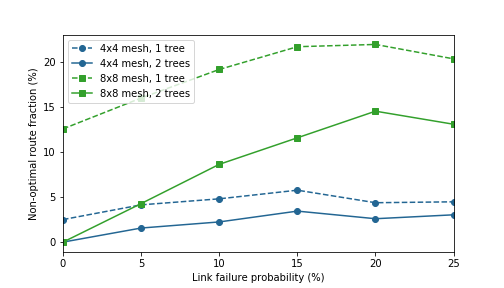}
    \caption{Fraction of source-destination pairs for which the chosen route is non-minimal, using 1 or 2 trees, in a 4x4 or 8x8 mesh network}
    \label{fig:non_opt_fraction}
\end{figure}

Similar results are observed when the fraction of source-destination pairs for which the returned route is always minimal is visualised. For more than 75\% of source-destination pairs the algorithm always returns paths with an optimal length. These results are shown in Fig. \ref{fig:non_opt_fraction}.

\begin{figure}[ht]
    \centering
    \includegraphics[width=0.41\textwidth]{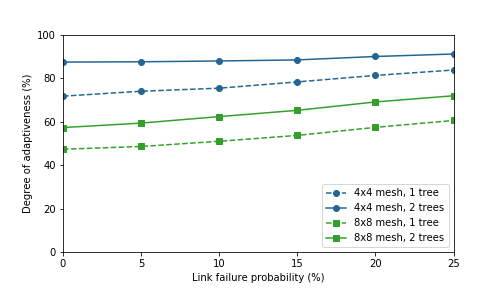}
    \caption{Average degree of adaptiveness for minimal length routes, using 1 or 2 trees, in a 4x4 or 8x8 mesh network}
    \label{fig:adaptiveness}
\end{figure}

Lastly, we evaluate the adaptiveness of the routing algorithm from a graph-based perspective. Provided a source-destination pair for which the routing algorithm exclusively finds paths with a minimal length, the degree of adaptiveness is defined as the number of legal routes which may be returned by the algorithm, divided by the number of minimal length paths in the graph \cite{Catania2006}. The average degree of adaptiveness for various setting is shown in Fig. \ref{fig:adaptiveness}. Once again, we see that a better adaptiveness is achieved when routing in a 4x4 mesh in comparison to the 8x8 mesh, and using a multi-tree algorithm improves the adaptiveness. As the occurrence of failures increases, and the number of minimal length paths drops, the results of our algorithm improve.

\section{Conclusion}  \label{sec:conclusion}
We introduced a topology-agnostic, fault tolerant, greedy, tree-based routing algorithm for networks-on-chip. The algorithm is fully distributed, and we have proven that it is deadlock free, without requiring any virtual channels.
Practical guidelines and optimisations for deploying the algorithm in a (hierarchical) grid network were provided. We executed simulations in 8x8 and 4x4 mesh network topologies to evaluate the quality of the provided routes and found that the generated routes were optimal for over 75\% of the source-destination pairs. The average ratio of the length of a generated path versus the length of a minimal path is below 1.14.

\bibliographystyle{IEEEtran}
\bibliography{bibliography}

\begin{thebibliography}{10}
\providecommand{\url}[1]{#1}
\csname url@samestyle\endcsname
\providecommand{\newblock}{\relax}
\providecommand{\bibinfo}[2]{#2}
\providecommand{\BIBentrySTDinterwordspacing}{\spaceskip=0pt\relax}
\providecommand{\BIBentryALTinterwordstretchfactor}{4}
\providecommand{\BIBentryALTinterwordspacing}{\spaceskip=\fontdimen2\font plus
\BIBentryALTinterwordstretchfactor\fontdimen3\font minus
  \fontdimen4\font\relax}
\providecommand{\BIBforeignlanguage}[2]{{%
\expandafter\ifx\csname l@#1\endcsname\relax
\typeout{** WARNING: IEEEtran.bst: No hyphenation pattern has been}%
\typeout{** loaded for the language `#1'. Using the pattern for}%
\typeout{** the default language instead.}%
\else
\language=\csname l@#1\endcsname
\fi
#2}}
\providecommand{\BIBdecl}{\relax}
\BIBdecl

\bibitem{Dally2001}
W.~J. Dally and B.~Towles, ``Route packets, not wires: on-chip interconnection
  networks,'' in \emph{Proceedings of the 38th Design Automation Conference
  (IEEE Cat. No.01CH37232)}, June 2001, pp. 684--689.

\bibitem{Constantinides2007}
K.~Constantinides, O.~Mutlu, T.~Austin, and V.~Bertacco, ``Software-based
  online detection of hardware defects mechanisms, architectural support, and
  evaluation,'' in \emph{40th Annual IEEE/ACM International Symposium on
  Microarchitecture (MICRO 2007)}, Dec 2007, pp. 97--108.

\bibitem{Abadal2017}
S.~Abadal, C.~Liaskos, A.~Tsioliaridou, S.~Ioannidis, A.~Pitsillides,
  J.~Solé-Pareta, E.~Alarcón, and A.~Cabellos-Aparicio, ``Computing and
  communications for the software-defined metamaterial paradigm: A context
  analysis,'' \emph{IEEE Access}, vol.~5, pp. 6225--6235, 2017.

\bibitem{Erika2012}
E.~Cota, A.~Amory, and M.~Lubaszewski, \emph{Reliability, Availability and
  Serviceability of Networks-on-Chip}.\hskip 1em plus 0.5em minus 0.4em\relax
  Boston, MA, USA: Springer US, 2012.

\bibitem{Chen2012}
L.~Chen and T.~M. Pinkston, ``Nord: Node-router decoupling for effective
  power-gating of on-chip routers,'' in \emph{2012 45th Annual IEEE/ACM
  International Symposium on Microarchitecture}, Dec 2012, pp. 270--281.

\bibitem{Murray2014}
J.~Murray, R.~Kim, P.~Wettin, P.~P. Pande, and B.~Shirazi, ``Performance
  evaluation of congestion-aware routing with dvfs on a millimeter-wave
  small-world wireless noc,'' \emph{J. Emerg. Technol. Comput. Syst.}, vol.~11,
  no.~2, pp. 17:1--17:22, Nov. 2014.

\bibitem{Radetzki2013}
M.~Radetzki, C.~Feng, X.~Zhao, and A.~Jantsch, ``Methods for fault tolerance in
  networks-on-chip,'' \emph{ACM Computing Surveys}, vol.~46, no.~1, pp.
  8:1--8:38, Oct. 2013.

\bibitem{Fattah2015}
M.~Fattah, A.~Airola, R.~Ausavarungnirun, N.~Mirzaei, P.~Liljeberg, J.~Plosila,
  S.~Mohammadi, T.~Pahikkala, O.~Mutlu, and H.~Tenhunen, ``A low-overhead,
  fully-distributed, guaranteed-delivery routing algorithm for faulty
  network-on-chips,'' in \emph{Proceedings of the 9th International Symposium
  on Networks-on-Chip}, ser. NOCS '15.\hskip 1em plus 0.5em minus 0.4em\relax
  New York, NY, USA: ACM, 2015, pp. 18:1--18:8.

\bibitem{Fukushima2009}
Y.~Fukushima, M.~Fukushi, and S.~Horiguchi, ``Fault-tolerant routing algorithm
  for network on chip without virtual channels,'' in \emph{2009 24th IEEE
  International Symposium on Defect and Fault Tolerance in VLSI Systems}, Oct
  2009, pp. 313--321.

\bibitem{Jovanovic2009}
S.~Jovanovic, C.~Tanougast, S.~Weber, and C.~Bobda, ``A new deadlock-free
  fault-tolerant routing algorithm for noc interconnections,'' in \emph{2009
  International Conference on Field Programmable Logic and Applications}, Aug
  2009, pp. 326--331.

\bibitem{Flich2012}
J.~Flich, T.~Skeie, A.~Mejia, O.~Lysne, P.~Lopez, A.~Robles, J.~Duato,
  M.~Koibuchi, T.~Rokicki, and J.~C. Sancho, ``A survey and evaluation of
  topology-agnostic deterministic routing algorithms,'' \emph{IEEE Transactions
  on Parallel and Distributed Systems}, vol.~23, no.~3, pp. 405--425, March
  2012.

\bibitem{Schroeder1991}
M.~D. Schroeder, A.~D. Birrell, M.~Burrows, H.~Murray, R.~M. Needham, T.~L.
  Rodeheffer, E.~H. Satterthwaite, and C.~P. Thacker, ``Autonet: a high-speed,
  self-configuring local area network using point-to-point links,'' \emph{IEEE
  Journal on Selected Areas in Communications}, vol.~9, no.~8, pp. 1318--1335,
  Oct 1991.

\bibitem{Jouraku2007}
A.~Jouraku, M.~Koibuchi, and H.~Amano, ``An effective design of deadlock-free
  routing algorithms based on 2d turn model for irregular networks,''
  \emph{IEEE Transactions on Parallel and Distributed Systems}, vol.~18, no.~3,
  pp. 320--333, March 2007.

\bibitem{Lysne2006}
O.~Lysne, T.~Skeie, S.~. Reinemo, and I.~Theiss, ``Layered routing in irregular
  networks,'' \emph{IEEE Transactions on Parallel and Distributed Systems},
  vol.~17, no.~1, pp. 51--65, Jan 2006.

\bibitem{Flich2002}
J.~Flich, P.~L{\'o}pez, J.~C. Sancho, A.~Robles, and J.~Duato, ``Improving
  infiniband routing through multiple virtual networks,'' in \emph{High
  Performance Computing}, H.~P. Zima, K.~Joe, M.~Sato, Y.~Seo, and
  M.~Shimasaki, Eds.\hskip 1em plus 0.5em minus 0.4em\relax Berlin, Heidelberg:
  Springer Berlin Heidelberg, 2002, pp. 49--63.

\bibitem{Ingebjorg2003}
I.~Theiss and O.~Lysne, ``Froots -- fault handling in up*/down* routed networks
  with multiple roots,'' in \emph{High Performance Computing - HiPC 2003},
  T.~M. Pinkston and V.~K. Prasanna, Eds.\hskip 1em plus 0.5em minus
  0.4em\relax Berlin, Heidelberg: Springer Berlin Heidelberg, 2003, pp.
  106--117.

\bibitem{Li2006}
M.~Li, Q.-A. Zeng, and W.-B. Jone, ``Dyxy - a proximity congestion-aware
  deadlock-free dynamic routing method for network on chip,'' in \emph{2006
  43rd ACM/IEEE Design Automation Conference}, July 2006, pp. 849--852.

\bibitem{Glass1992}
C.~J. Glass and L.~M. Ni, ``The turn model for adaptive routing,'' in
  \emph{[1992] Proceedings the 19th Annual International Symposium on Computer
  Architecture}, May 1992, pp. 278--287.

\bibitem{Chiu2000}
G.-M. Chiu, ``The odd-even turn model for adaptive routing,'' \emph{IEEE
  Transactions on Parallel and Distributed Systems}, vol.~11, no.~7, pp.
  729--738, July 2000.

\bibitem{Sahhaf2013}
S.~Sahhaf, W.~Tavernier, D.~Colle, M.~Pickavet, and P.~Demeester, ``Link
  failure recovery technique for greedy routing in the hyperbolic plane,''
  \emph{Computer Communications}, vol.~36, no.~6, pp. 698 -- 707, 2013,
  reliable Network-based Services.

\bibitem{Bjerregaard2006}
T.~Bjerregaard and S.~Mahadevan, ``A survey of research and practices of
  network-on-chip,'' \emph{ACM Comput. Surv.}, vol.~38, no.~1, Jun. 2006.

\bibitem{Catania2006}
V.~Catania, R.~Holsmark, S.~Kumar, and M.~Palesi, ``A methodology for design of
  application specific deadlock-free routing algorithms for noc systems,'' in
  \emph{Proceedings of the 4th International Conference on Hardware/Software
  Codesign and System Synthesis (CODES+ISSS '06)}, Oct 2006, pp. 142--147.

\end{thebibliography}

\end{document}